\documentclass[journal=aamick,manuscript=letter,final]{achemso}

\usepackage[version=3]{mhchem} 

\usepackage{color}

\author{Soumen Mandal}
\email{mandals2@cardiff.ac.uk, soumen.mandal@gmail.com} \affiliation[Cardiff University]{School of Physics and Astronomy, Cardiff University, Cardiff, UK}
\author{Jerome Cuenca}
\affiliation[Cardiff University]{School of Physics and Astronomy, Cardiff University, Cardiff, UK}
\author{Fabien Massabuau}
\affiliation[Cambridge]{Department of Materials Science \& Metallurgy, University of Cambridge, Cambridge, UK}
\author{Chao Yuan}
\affiliation[Bristol]{Center for Device Thermography and Reliability, Bristol University, Bristol, UK}
\author{Henry Bland}
\affiliation[Cardiff University]{School of Physics and Astronomy, Cardiff University, Cardiff, UK}
\author{James W. Pomeroy}
\affiliation[Bristol]{Center for Device Thermography and Reliability, Bristol University, Bristol, UK}
\author{David Wallis}
\affiliation[Cardiff engin]{School of Engineering, Cardiff University, Cardiff, UK}
\alsoaffiliation[Cambridge]{Department of Materials Science \& Metallurgy, University of Cambridge, Cambridge, UK}
\author{Tim Batten}
\affiliation{Renishaw plc., Wotton-under-Edge, UK}
\author{David Morgan}
\affiliation{Cardiff Catalysis Institute, School of Chemistry, Cardiff University,  Cardiff, UK}
\author{Rachel Oliver}
\affiliation[Cambridge]{Department of Materials Science \& Metallurgy, University of Cambridge, Cambridge, UK}
\author{Martin Kuball}
\affiliation[Bristol]{Center for Device Thermography and Reliability, Bristol University, Bristol, UK}
\author{Oliver A. Williams}
\email{williamso@cardiff.ac.uk}
\affiliation[Cardiff University]{School of Physics and Astronomy, Cardiff University, Cardiff, UK}

\title{Thick adherent diamond films on AlN with low thermal barrier resistance}
\begin{document}

\begin{abstract}
Growth of $>$100 $\mu$m thick diamond layer adherent on aluminium nitride is presented in this work. While thick films failed to adhere on untreated AlN films, hydrogen/nitrogen plasma treated AlN films retained the thick diamond layers. Clear differences in zeta potential measurement confirms the surface modification due to hydrogen/nitrogen plasma treatment. Areal Raman maps showed an increase in non-diamond carbon in the initial layers of diamond grown on pre-treated AlN. The presence of non- diamond carbon has minimal effect on the interface between diamond and AlN. The surfaces studied with x-ray photoelectron spectroscopy (XPS) revealed a clear distinction between pre-treated and untreated samples. The surface aluminium goes from nitrogen rich environment to an oxygen rich environment after pre-treatment. Cross section transmission electron microscopy shows a clean interface between diamond and AlN. Thermal barrier resistance between diamond and AlN was found to be in the range of 16 m$^2$K/GW which is a large improvement on the current state-of-the-art.  \end{abstract}

\section{Introduction}
High electron mobility transistors (HEMTs) made from gallium nitride (GaN) are current state-of-the-art for high power and high frequency applications\cite{mishra2002, mishra2008}. Still, the full potential of GaN HEMT devices is yet to be realised due to less than effective thermal management during high power and high frequency operations. Lee et al.\cite{lee2008} have shown that slight changes to operating temperature of these devices can lead to significant reduction in device lifetime. Current standards for effective thermal management in GaN devices entails the structures to be made from GaN grown on silicon carbide (SiC)\cite{gaska1998, kuball2002}. The high thermal conductivity of SiC ($\kappa_{SiC} \approx$ 360-490 W /m K \cite{goldberg2001}) has resulted in good device performances, but, further improvement can be possible if the SiC layer is replaced with a diamond layer ($\kappa_{SiC} \approx$ 2100 W/m K \cite{ho1972} for single crystal and $\approx$ 1200 W/m K\cite{coe2000, iva2002, pome2014} for polycrystalline diamond). For cost effective implementation of diamond as thermal management layer polycrystalline films need to be used. This is due to the fact that it is possible to grow large area polycrystalline diamond films which is not the case for single crystal. Furthermore, it has been shown that the large thermal conductivity of diamond (greater than SiC to improve device performance) is possible in layers which are more than 100$\mu$m thick\cite{ana2016}.

For fabrication of GaN-diamond devices, the diamond can be grown on GaN or GaN can be grown on diamond. The growth of GaN on diamond is possible only on single crystal diamond and has led to promising devices\cite{jess2006, hira2011}. But, the high cost and small sizes of single crystal diamond makes them unattractive. In the past polycrystalline diamond has been grown on adhesion layer buffered GaN with limited success\cite{dum2013, pome2014}. This is due to the poor thermal conductivity of the amorphous adhesion layer\cite{sun2015}. Alternatively, we have shown the growth of thin diamond layers directly on GaN surface\cite{man2017}. Unfortunately, it is not possible to grow thick diamond layer ($>$10 $\mu$m) on GaN due to absence of any carbide bond between GaN and diamond layers. It has also been shown that GaN grown on sapphire is unsuitable for  growth of thick diamond layers due to thermal mismatch between diamond and sapphire\cite{may1994}. Alternatively, GaN can also be grown on silicon but with aluminium nitride (AlN) as a buffer layer\cite{zhu2013}. While the devices will be fabricated on the GaN side, the diamond can be grown on the AlN side by transferring the wafer to carrier wafer and exposing the AlN side. By careful etching of the AlN side it is possible to etch all the way to the single crystal AlN layer buried deep in the GaN-AlN stack. 

In this work we have detailed the growth of thick ($>100 \mu$m) diamond layer on 250nm thick AlN layer. The thermal barrier resistance of the interface was also measured. In the past there have been attempts to grow thin diamond layer on AlN\cite{god1995, cui1996, wang2000, hees2013, cerv2012} but the growth of $>$50$\mu$m thick layers have not been attempted. The growth of diamond on AlN surface by microwave plasma chemical vapour deposition also exposes the surface to hydrogen and methane plasma. Pobedinskas et al.\cite{pobe2014} had studied the effects of hydrogen and methane plasmas on sputtered AlN. They found that the films can be etched when exposed to plasma for long periods of time. That means the AlN layer has to be exposed to the plasma for as little time as possible. Once the diamond film is coalesced, the AlN layer should be protected by the diamond layer. Cervenka et al.\cite{cerv2013} had studied the effect of hydrogen plasma on nucleation density  and surface morphology of diamond grown on single crystal AlN. Here, we have measured the $\zeta$ - potential of the AlN films as a function of pH. The growth of thick layers on as-received AlN was not possible but after surface treatment it was possible to grow thick layers on AlN thin films. We have also measured the $\zeta$ - potential  of the AlN surfaces after treatment. X-ray photoelectron spectroscopy was done to analyse the surfaces of treated and untreated surfaces. Atomic force microscopy (AFM) was used to observe the diamond seeds on the surface of the substrates. We have performed Raman maps on thin diamond films grown on as-received and pre-treated AlN to determine the interface composition. The interface was imaged and electron energy loss spectroscopy was done using Transmission Electron Microscopy. Finally the thermal barrier resistance between diamond and AlN was measured using transient thermoreflectance measurements.

\section{Methods}
The work in this study was carried out on AlN films grown on Si substrate. The AlN layers were grown on to 150mm Si substrates in an Aixtron close coupled showhead Metal organic chemical vapour deposition (MOCVD) system. Trymethyl Aluminium (TMAl) was used as the Al source and ammonia (NH3) was used as the Nitrogen source. The AlN layer was approximately 250nm thick and had a Al-polar orientation.

\subsection{Zeta potential measurement}
The $\zeta$-potential of the substrates (AlN on Si) were measured by measuring the streaming potential of the substrates in Surpass$^{TM}$ 3 electrokinetic analyzer. The change in potential or current between two Ag/AgCl electrodes at the either end of streaming channel as a function of electrolyte pressure gives the streaming potential. The shearing of the counterions in the streaming channel by flowing electrolytes from the charged surface gives rise to a streaming current/potential between two electrodes positioned tangentially to the flowing electrolyte. The flow of counterions in the channel is proportional to the electric double layer of the surfaces forming the channel. Hence, the measured streaming current/voltage is related to the $\zeta$ - potential of the surface\cite{wag1980}. The streaming potential technique is not only useful for well defined surfaces but also can be used for measuring $\zeta$ - potential of fibres\cite{jacob1985}. The setup for measuring $\zeta$ - potential was suggested by Van Wagenen et al.\cite{wag1980} and has been used for measuring $\zeta$ - potential of variety of flat surfaces\cite{voi1983, nor1990, sca1990, man2017, bla2019, man2019}. In our experiments, the streaming channel was formed by two plates of AlN on Si. The channel width was kept between 90-110 $\mu$m. A 10$^{-3}$ M solution of KCl in DI water was used as an electrolyte. The pressure of the electrolyte was varied between 600 and 200 mbar. For altering the pH of electrolyte, 0.1M NaOH and 0.1M HCl solution was used with the inbuilt titrator in Surpass$^{TM}$ 3.

\subsection{Diamond growth}
The wafers were seeded with mono-dispersed diamond solution. The growth of diamond on the seeded wafers were done using a Carat Systems SDS6U microwave chemical vapour deposition systems. 15 X 15 mm$^2$ pieces of AlN on Si were seeded with both H - and O - terminated seeded and placed in the reactor. A gas mixture of 3\% CH$_4$/H$_2$ at 110 torr with 5kW microwave power was used for the growth. A thick diamond layer ($>$100$\mu$m) was grown over a time period of 48 hours. After growth the samples were cooled down in the growth mixture over a 2 hour time period. For both H- and O- terminated seeds complete diamond layers were formed after 48 hours growth. The grown samples were then laser cut to get rid of the edges sticking to the silicon substrate underneath the AlN layer. After laser cutting it was found the diamond layer on AlN immediately delaminated. This may be due to excessive stress in case of H - terminated seeds and inadequate adhesion in the case of O- terminated seeds. In the literature, it has been shown\cite{cho2000} that as-grown AlN has large number of dangling bonds. A treatment with nitrogen plasma helps to stabilise the AlN surface and reduce the dangling bond density. In our case we used a 10\% N$_2$/H$_2$ gas mixture at 1.5 kW microwave plasma and 20 torr to pretreat the surface before seeding. The treatment times were fixed at 5, 10 and 15mins. Since the H- terminated seeds failed to produce diamond layers which did not delaminate only O- terminated seeds were used on the plasma treated samples. The substrates were seeded after plasma treatment and a thick diamond layer was grown. The samples were again laser cut and it was found that diamond layers stayed attached to AlN only when the pretreatment time was 10minutes or more. So, for the rest of the study we have used only 10 minutes plasma treatment before seeding. 

 \subsection{X-Ray Photoelectron Spectroscopy}
 X-ray photoelectron spectroscopy (XPS) was used to study the surface of the as-received and plasma treated substrates. A Thermo Scientific K-Alpha$^+$ spectrometer equipped with a monochromatic Al source operating at 72W (6 mA emission current at 12kV anode potential) was used to collect the spectra. Pass energies of 150 and 40 eV were used for acquiring survey and high-resolution spectra respectively. For charge neutralisation a combination of electrons and low energy Ar ions. The spectra were analysed using CasaXPS software. XPS spectra of both treated and untreated surfaces were acquired. 
 
 \subsection{Atomic Force Micrscopy}
The seeding density was checked using a Veeco Dimension Icon with SCANASYST-AIR (Bruker-AFM probes supplied, 2nm nominal radius) tips operated in peak force tapping mode. The analysis of the images were done using Gwyddion\cite{necas2012} scanning probe microscopy analysis software.

\subsection{Raman Spectroscopy}
 We have done areal Raman maps of 50nm diamond films grown on treated and untreated AlN films. The measurements were done using a Renishaw InVia Qontor Raman spectrometer was used in backscatter geometry, equipped with a 532 nm laser excitation wavelength, 2400 l/mm grating and a 100x lens (N.A. = 0.85). The maps were taken over 30 $\times$ 30 $\mu$m area with a step size of 1 $\mu$m. The spectral resolution was 1.5 cm$^{-1}$. The range of measurement was covered by taking two separate high resolution scans over the same area. The scans were taken between 400-1600 and 1230-2300 cm${-1}$. 

\subsection{Cross-sectional TEM}
We have carried out the atomic and chemical characterisation of the AlN-diamond interface  using transmission electron microscopy (TEM). The samples were prepared using focused ion beam (FIB). An aberration-corrected FEI Titan3 operated at 300 kV was used for high angle annular dark field scanning TEM (HAADF-STEM) imaging, and an FEI Tecnai Osiris operated at 200 kV was used for electron energy loss spectroscopy (EELS).

\subsection{Thermal Properties}

Thermal properties of AlN/Diamond films (Si substrate removed) were characterised using a transient thermoreflectance technique (TTR). A 10nm Cr adhesion layer was deposited onto the surface of the AlN followed by a 100nm thick Au layer to act as a transducer during TTR measurements. A 355 nm heating pump beam with a duration of 1ns, was used; induced surface temperature changes were probed with a continuous wave 532 nm laser. The pump beam had a Gaussian profile (1/e$^2$ radius of 85 $\mu$m), the probe beam a diameter of $<$2$\mu$m, located in the centre of pump spot. A temperature transient model fitted to the measurement was used to determine the unknown thermal properties in multilayer material stack (Au/Cr-AlN-Diamond in this study), with more details described in\cite{tou2001, gar2015, zho2017}. The fitted thermal properties include the thermal interfacial resistance at the Au-AlN interface (R$_{\mbox{\tiny{Au-AlN}}}$), the thermal resistance across AlN layer and AlN-diamond interface (R$_{\mbox{\tiny{AlN-Di}}}$) and the diamond thermal conductivity (k$_{\mbox{\tiny{di}}}$).


\section{Results and Discussions}
\subsection{Zeta potential measurements}
\begin{figure}\centerline{\includegraphics[height=3in,angle=0]{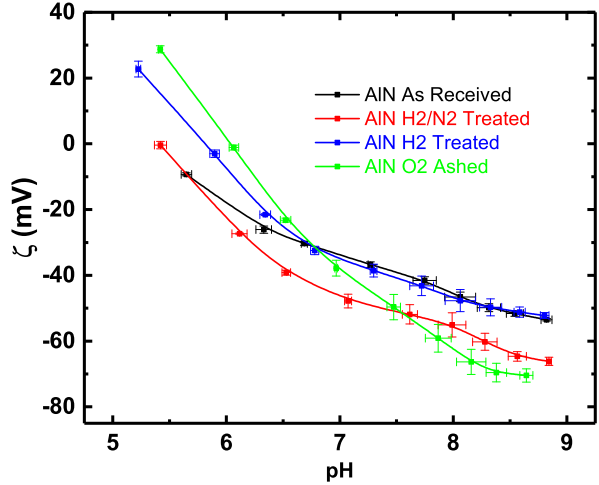}}
\caption{Zeta potential vs pH for as-received AlN and the same sample after pretreatment with H$_2$/N$_2$ plasma for 10mins. A clear increase in negative zeta potential can be seen after treatment indicating an increase in oxygen content on the surface. The zeta potential of the AlN surface after H$_2$ plasma treatment and oxygen ashing is also presented. The change in zeta potential for these treatments are not as large as for H$_2$/N$_2$ plasma at pH 7.} \label{zeta}
\end{figure}

The results of the $\zeta$-potential measurements are shown in figure \ref{zeta}. To start with we measured the $\zeta$-potential of as-received AlN film. The films were then seeded with both H and O- terminated seeds. The H-terminated seeds are known to have positive $\zeta$-potential while the O-terminated are negatively charged in water\cite{hee2011}.  Even though, $\zeta$-potential studies point towards H- terminated seeds (positively charged seeds on negatively charged surface), O-terminated seeds were also used to cross-check validity of the measurements. AFM measurements performed on the seeded wafers reveal seeding densities in accordance with $\zeta$-potential values. The thick diamond films grown on as-received AlN delaminated immediately after growth. Hence, a pre-treatment technique was used for further experiments. The diamond grown on pre-treated AlN was tested with Raman spectroscopy and Scanning electron microscopy for quality. 

We have measured $\zeta$-potentials of the as-received, H$_2$/N$_2$ plasma, H$_2$ plasma and O$_2$ plasma treated AlN substrates as a function of electrolyte pH. The main point of interest for seeding is the $\zeta$-potential in the pH range of 6-7. This is the pH range of the diamond solution. In this range the potentials for H$_2$ plasma and O$_2$ plasma treated samples remain close to the as-received substrates, staying in the range of -30 to -35mV. In contrast the $\zeta$ - potential of H$_2$/N$_2$ plasma treated substrates is enhanced and it is between -40 to -45 mV in the pH range of interest.  Since, an observable change in $\zeta$ - potential was only seen in case of H$_2$/N$_2$ plasma, only samples pretreated with this recipe were extensively studied. 

\subsection{X-ray photoelectron spectroscopy}
\begin{figure}\centerline{\includegraphics[width =5in,angle=0]{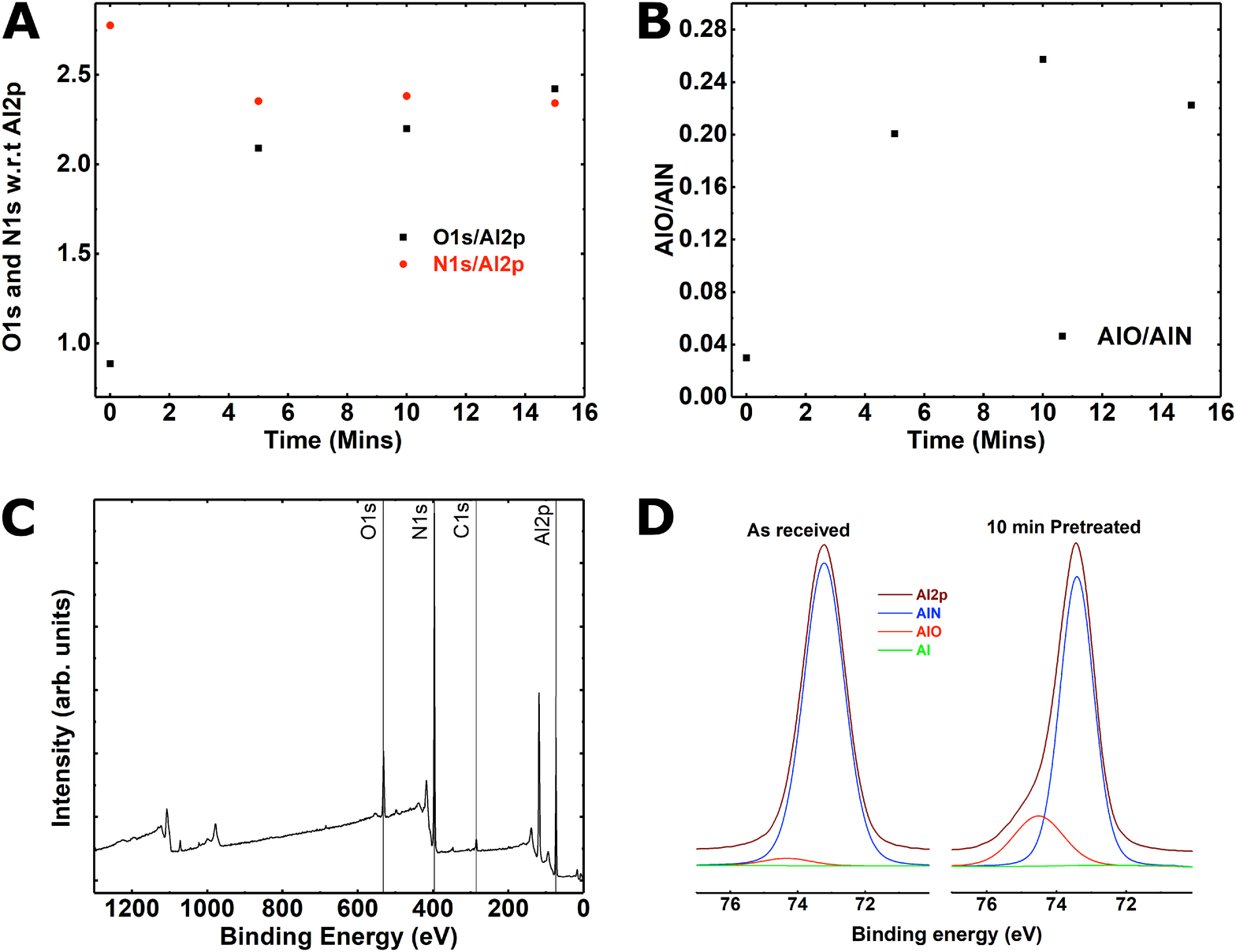}}
\caption{A: Shows the ratio of O1s and N1s to Al2p on the surface of AlN film before after plasma treatment. B: The ratio of AlO to AlN peaks in the XPS data has been shown here. The peaks have been determined by deconvoluting the Al2p peak in the spectra. C: A survey spectra for an AlN sample is shown here. The vertical lines mark the positions of peaks of interest. D: Al2p peak for as-received and 10min treated samples are shown here. The observed data has been shifted by a fixed amount to show the individual fitted (AlN, AlO and Al) peaks.} \label{xps}
\end{figure}
Figure \ref{xps} shows the ratios of various species on the surface of AlN films before and after pre-treatment. For the analysis of the spectra the charge correction was done with respect to the adventitious carbon peak at 285 eV\cite{wilson2001, ferro2005}. After the correction the O1s peak was found to be at 531.2 $\pm$ 0.4 eV. The data was taken on both as-received as well as samples that were cluster argon etched in-situ to get rid of surface contaminants.  In panels A and B of figure \ref{xps} we have shown the data for cluster etched samples only. In panel C of the same figure we have shown the survey spectra of one of the samples. The vertical lines have been drawn to mark the positions of Al2p, C1s, N1s and O1s peaks. The zoomed in Al2p peak from the as-received and 10min treated samples are shown in figure \ref{xps}D. The observed spectra has been purposely shifted to show the individual fitted curves.

For deconvoluting the Al2p peak three components were chosen, AlN, AlO and Al. While none of the samples showed presence of Al on the surface(green curve in panel figure \ref{xps}D) the AlN and AlO species had varying concentrations.  There are a wide ranging of values available for such peaks in literature\cite{youngman1990, liao1993, kazan2005, kim2009, jose2010, garcia2011, alevli2012, motamedi2014}, we have taken values from Alevli et al. \cite{alevli2012} as our starting point.  Based on this, we have assigned certain constraints to the values of AlO and Al peaks with respect to AlN peak. The AlO peak was constrained to be at +1eV from AlN peak and the Al peak was constrained to be at -0.9eV from the same peak. All the four samples were analysed using this method both before and after cluster etching. The peak position for AlN was found to be 73.3 $\pm$ 0.3 eV and that for AlO was 74.5 $\pm$ 0.3 eV which is in good agreement with published data. 

To see the effects of pretreatment we compare the total intensities of various peaks in the XPS data. In figure \ref{xps}A we have compared the O1s and N1s peak to Al2p peak. While the N1s to Al2p ratio goes down the ratio between O1s to Al2p goes up. This is a clear indication that the surface Al goes from a nitrogen rich environment to a more oxygen rich environment. To see the relative concentrations of AlO and AlN we have compared the two in figure \ref{xps}B. It is clear that the surface is predominantly nitrogen rich but after the treatment the surface has almost 25\% oxygen. The increase in the oxygen content at the surface is also validated by decrease in zeta potential of AlN films before and after treatment. 

\subsection{Atomic Force Microscopy}
The substrates were seeded with monodispersed diamond/H$_2$O solution in an ultrasonic bath. This type of seeding produces nucleation density in excess of 10$^{11}$ cm$^{-2}$ on silicon wafers\cite{olie2007seed}. The $\zeta$ - potential of seed solution is dependent on the surface termination of the diamond particles. H - terminated diamond particles give rise to positive $\zeta$ potential and O - terminated seeds lead to negative potential\cite{olie2010hyd}. Commercially available nanodiamond seeds are aggregated and oxygen terminated\cite{mochalin2011}. The procedure for creating diamond solutions for seeding has been described elsewhere\cite{olie2010hyd}. We have treated the as-received and H$_2$/N$_2$ plasma treated wafers with both H - and O - terminated diamond seed solution. 

\begin{figure}\centerline{\includegraphics[width=4in,angle=0]{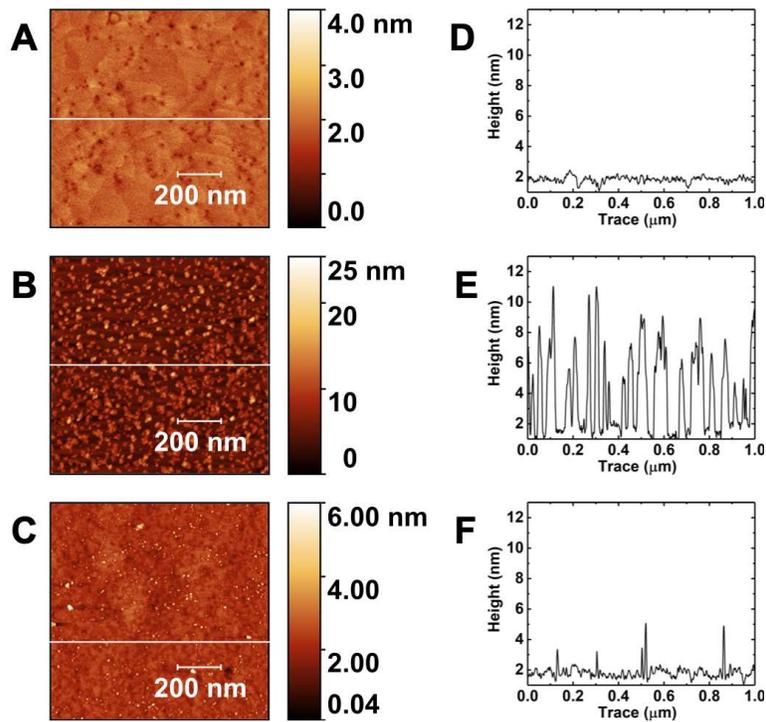}}
\caption{AFM images of AlN wafers before (Panel A) and after seeding with hydrogen (Panel B) and oxygen(Panel C) terminated diamond seed solution. The wafer was plasma pre-treated before seeding.  Panels D, E and F are the line profiles of the left column images as indicated by the white line.} \label{afm}
\end{figure}

From the $\zeta$-potential results it is clear that we have to use H-terminated diamond seeds, which are positively charged, to seed the AlN substrate. We have seeded the surfaces with both H - and O - terminated seeds. Atomic force microscopy (AFM) was done on seeded and unseeded substrates after plasma pre-treatment. The micrographs have been plotted in figure \ref{afm}. Panel A is the surface after pretreatment.  For a scan area of 1 $\mu$m$^2$ the roughness values of the substrates, as calculated from AFM data,  are not effected by the pr-treatment. As expected, we see high seeding density for H- terminated seeds (Panel B). Panel C is the micrograph for a sample seeded with O-terminated seeds. Panels D , E and H are the line profiles of the AFM images as indicated be the white lines on the images. Panel D, E and F correspond to unseeded, H- and O- terminated seeded surfaces after pre-treatment respectively.  The line trace in panel E shows presence of large number of seeds on the surface. On the other hand, the line trace in panel F shows very little seeds. But one comparison with panel D we can confirm that there are some seeds on the surface. The seeding densities observed are in-line with $\zeta$-potential results.

\subsection{Raman Spectroscopy}
\begin{figure}\centerline{\includegraphics[height=5in,angle=0]{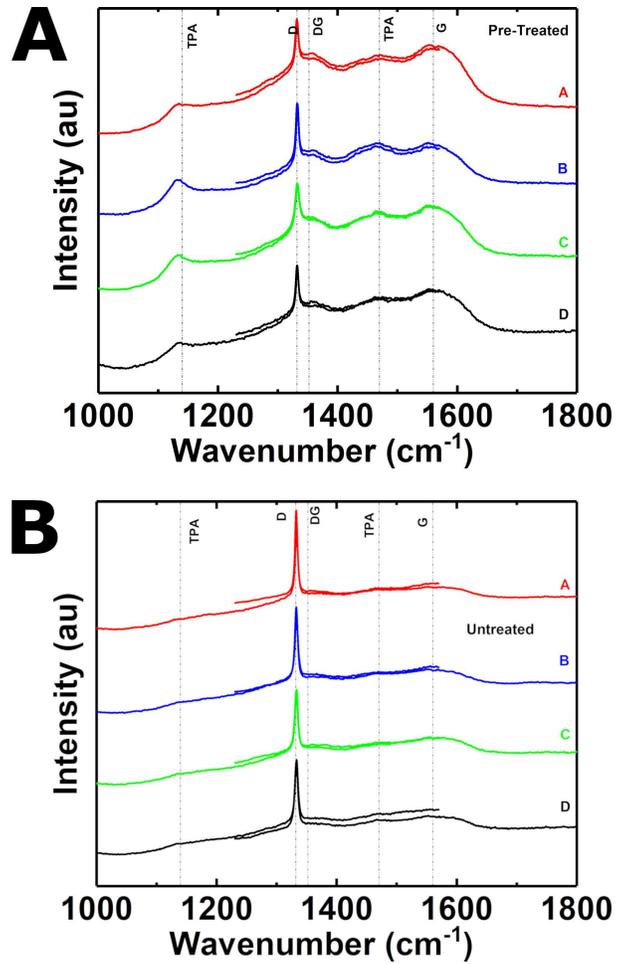}}
\caption{Raman spectroscopy data from different areas on very thin diamond grown on pre-treated and untreated AlN. The most common peaks have been marked on the figures.} \label{raman}
\end{figure}

We have acquired the Raman maps of two 50nm thin diamond films grown on pre-treated and untreated AlN films to understand the diamond-AlN interface. The spectrum was taken over an area of 30 $\times$ 30 $\mu$m.  The high-resolution scans were taken between 400-1600 and 1230-2300 cm$^{-1}$. The spectra from two different scans were joined by normalising the spectra with respect to 1332 cm$^{-1}$ diamond peak\cite{ram1930, bhag1930}. The pre-treatment in this case was only 10min H$_2$/N$_2$ plasma exposure before seeding. From these maps we have selected four random points and the Raman data from those points are presented in figure \ref{raman}. Panel A is for the sample grown on pre-treated AlN and panel B is for the diamond grown on untreated AlN. The standard peak positions of various components are marked in the graph.

The most possible bonding between the diamond and AlN layer can be a carbide bond or cyanide bond. The aluminium carbide peaks should appear at 718 and 864 cm$^{-1}$\cite{ken2015} and the cyanide peaks should appear as a band around 2090 cm$^{-1}$\cite{seki1993, cos2016}. In all our data we see no existence of any peaks or bands. This clearly means that no large scale bonding through formation of carbide or cyanide containing layers is occurring in these films. Looking at the figure it is evident that there is some remarkable difference between the non-diamond carbon content in the two thin films. We have marked the position of trans-polyacetylene (TPA) peaks at 1150 and 1450 cm$^{-1}$\cite{fer2001} along with D (DG) and G (G) at 1350 and 1560 cm$^{-1}$\cite{may2008} respectively. The disordered carbon peak at 1405 cm$^{-1}$, linked to graphitic rings, is not present in these thin films and the shift of G band from 1580 to 1560 cm$^{-1}$ is mainly due to switchover of the $\pi$ ring system to $\pi$ chain systems\cite{fer2004}.  All the major peaks related to non-diamond carbon, marked on the Raman spectra, are quite prominent in the film grown on pre-treated AlN when compared with film grown on untreated AlN. The large difference in the non-diamond carbon content in the initial phases of growth may be the key to the adhesion of the thick diamond layer on the AlN surface. We have further done cross-sectional scanning transmission electron microscopy (STEM) to characterise the film interface.

\subsection{Cross-sectional STEM}
\begin{figure}\centerline{\includegraphics[width=6in,angle=0]{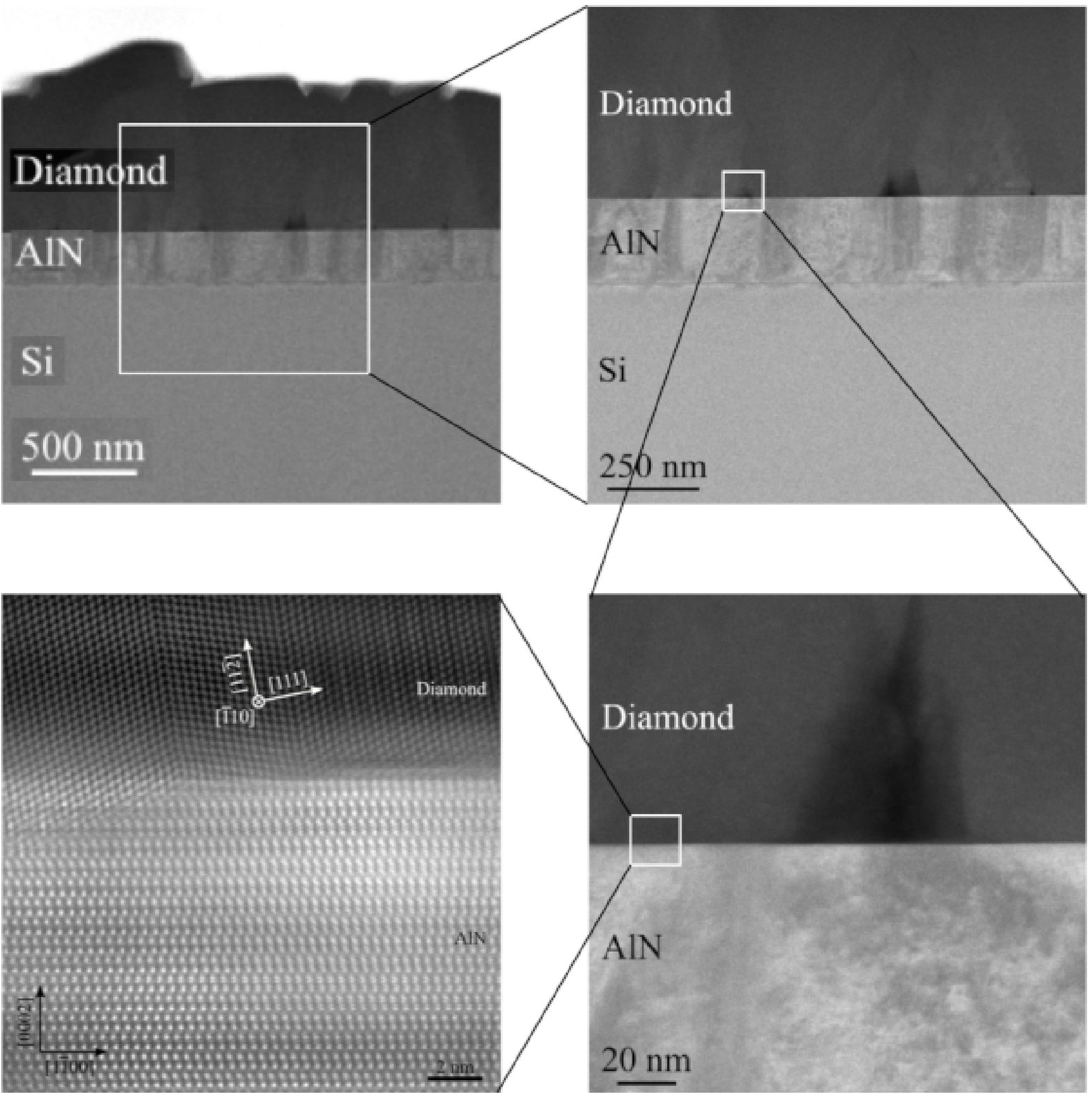}}
\caption{HAAD-STEM image of the sample grown using O-treated diamond seeds. Holes can be seen in places in the diamond film, originating from the seeding. High-resolution TEM images reveal that the interface is sharp with no obvious intermediate phase forming between the AlN and the diamond. It should be noted that the diamond film is polycrystalline and the structure of the interface may vary based on the orientation of the diamond seed relative to the substrate.} \label{csec}
\end{figure}
\begin{figure}\centerline{\includegraphics[width=5in,angle=0]{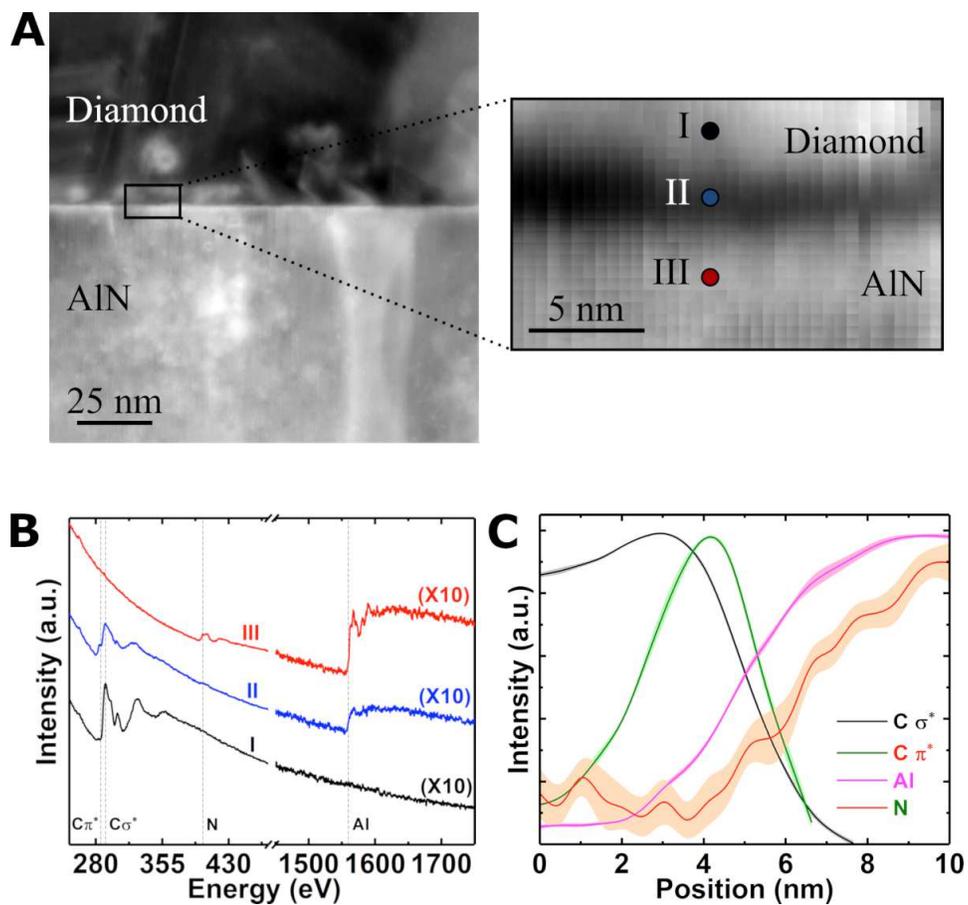}}
\caption{Panel A: Dark field image obtained in the same region as the high resolution image in figure 4 (main text) and region where the EELS data where recorded; B: Selected EELS spectra taken in the diamond film (I), at the interface (II) and in the AlN substrate (III) - the peaks of interest are labelled; C: Plot of the (normalised) intensity of the peaks of interest across the interface.  } \label{eels} \end{figure}

The AlN-diamond interface was observed in cross section by high angle annular dark field STEM (HAADF-STEM) and is shown in figure \ref{csec}. Starting from the top-left image the, the other images are subsequent zoom as indicated in the figure. Voids at the interface are clearly visible. This is because of the poor seeding density that arises due to the O-terminated seeds used in seeding. The bottom left image shows a clean interface with little or no evidence of any carbide or cyanide containing layers as seen from the Raman data as well.  This image of the interface is atomically resolved, and that is achieved in this region of the sample because the diamond and AlN lattices are locally well aligned.  This alignment is not, however, typical since the diamond film is polycrystalline and thus not uniformly aligned to the AlN lattice. The EELS, shown in figure \ref{eels},  taken at same place as the bottom left image, reveals the possible existence of atomically thin carbide layer, not detectable by Raman. This is inferred from the fact that, at the interface the Al K-edge is detected, but no N K-edge. This extremely thin layer may be responsible for high adhesion of the diamond films as well as the low thermal barrier as discussed in the next section. The possible carbide layer seen by EELS is not detected by Raman due it being atomically thin.

\subsection{Thermal properties}
\begin{figure}\centerline{\includegraphics[width=6in,angle=0]{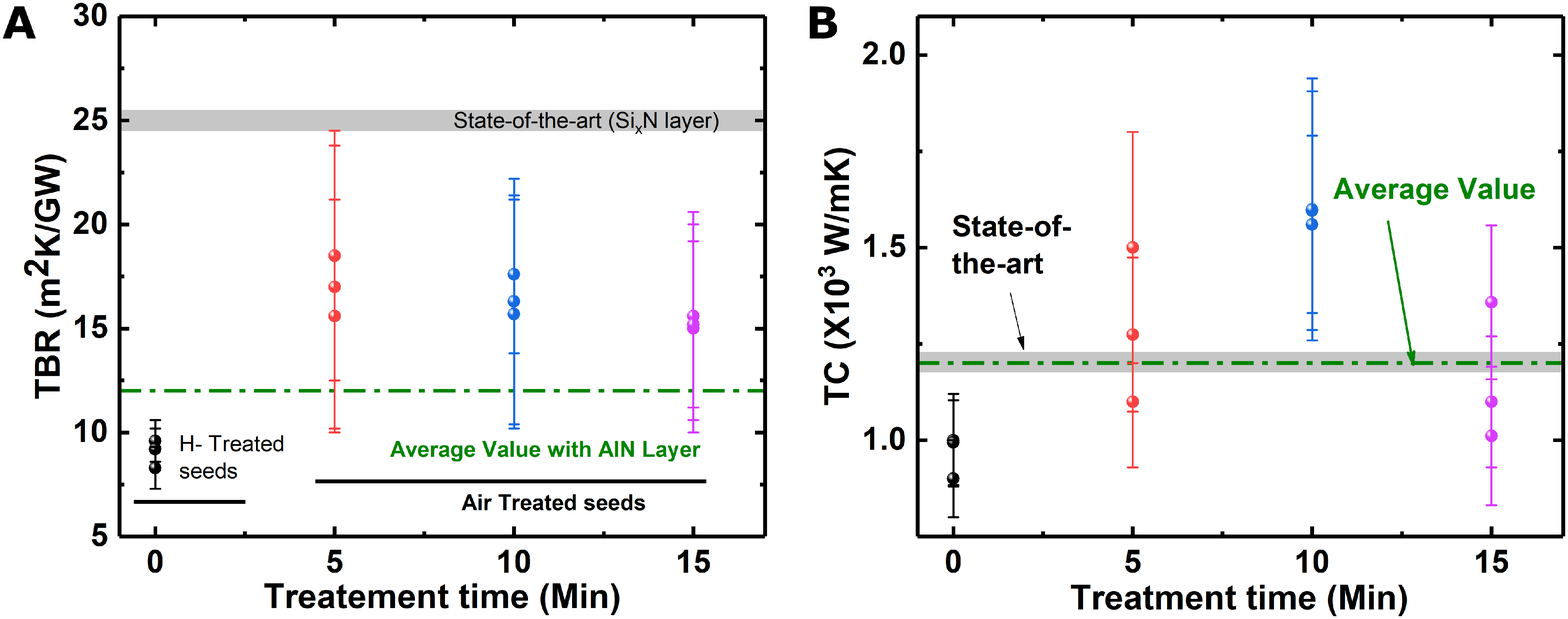}}
\caption{A: The thermal boundary resistance of AlN-diamond interface measured with TTR technique is shown here. We have measured four samples with varying pre-treatment time and seeds. B. Shows the thermal conductivity of the thick ($\sim$ 100 $\mu$m) diamond layer grown on AlN layer.} \label{tbr}
\end{figure}

We have measured the thermal properties of the AlN-diamond interface. Figure \ref{tbr}A shows the measured R$_{\mbox{\tiny{AlN-Di}}}$ benchmarked of samples studied in this work against the state-of-the art thermal boundary resistance (TBR$_{\mbox{\tiny{eff}}}$) data of GaN-on-Diamond using SiN as the interlayer\cite{sun2015}. The most relevant of all the samples is the one with 10 min treatment. The sample with hydrogen treated seeds show the lowest barrier resistance, but, as explained above it is not possible to get large scale thick ($>$ 50$\mu$m)intact films with hydrogen treated seeds. The diamond samples grown with AlN interlayer demonstrate the greatly reduced thermal resistance. The main reason is that the crystalline AlN seed layer grown by MOCVD is a higher thermal conductivity material. The effective thermal conductivity of AlN interlayer was estimated to be as high as 30 W/m.K, by the ratio of AlN thickness (250 nm) to  R$_{\mbox{\tiny{AlN-Di}}}$; this is in contrast to 1-3 W/m.K\cite{sun2015} for amorphous Si$_3$N$_4$ commonly used to the seed the diamond growth. We note that bulk single crystalline AlN can have a thermal conductivity as high as $\sim$400 W/m.K\cite{rou2018}; thermal conductivity of a 250 nm single-crystalline AlN layer in the sample can be estimated to be about 150 W/m.K using the modified Debye-Callaway model due to boundary scattering\cite{mor2002}, suggesting point defects or grain boundaries reduces its thermal conductivity somewhat. The high resolution TEM micrograph of the AlN-diamond interface shown in figure \ref{csec} illustrates that there will be only a minor amount of interface roughness phonon scattering at the AlN-diamond interface due to defects. We note that minimum thermal resistance TBR achievable at AlN/Diamond interface from a diffuse mismatch model (DMM), relying only upon the density of states in these two materials\cite{swa1989}, is 0.8 m$^2$K/GW.  

The average of measured k$_{\mbox{\tiny{di}}}$ results is $\sim$1200 W/m.K (shown in figure \ref{tbr}B) for the diamond of $\sim$100$\mu$m thickness. This comparable to results expected for polycrystalline diamond of similar thickness \cite{ana2016}. We note this diamond thermal conductivity is an average thermal conductivity of diamond through the whole layer thickness; with the smaller diamond crystalline size near the AlN-diamond interface thermal conductivity of the diamond near this interface is typically reduced \cite{ana2016}.

\section{Conclusions}
In conclusion we have shown that it is possible to grow thick ($>$100 $\mu$m) diamond layer on AlN. Such a layer would be used for thermal management of GaN high power devices. For successful growth, the AlN layers need to be pre-treated with 10\% N$_2$/H$_2$ plasma for 10 mins. The pre-treatment has shown to increase the oxygen content of the AlN surface, thus making its $\zeta$-potential more negative. Even though H-treated diamond seeds (positively charged seeds on negatively charged surface) seem to be obvious choice based on $\zeta$-potential study, it was found that films grown after seeding with such seeds delaminated very quickly. As a result, it was found that O-terminated seeds after pre-treatment resulted in films which did not delaminate. Raman spectroscopy studies revealed that the sp$_2$ carbon content in the initial stages of growth is much higher in films grown on pre-treated AlN. It is possible that the excess non-diamond carbon assists in the adhesion of the diamond to AlN layer. The diamond-AlN interface was found to be extremely abrupt. The average thermal barrier resistance measured for the samples grown with O-terminated seeds was found to be $\sim$16 m$^2$K/GW which is much lower than the current state-of-the-art.


\section*{Acknowledgements}
The authors would like to acknowledge financial support of the Engineering and Physical Sciences Research Council under the program Grant GaN-DaME (EP/P00945X/1). SM would like to thank Dr Andreas Papageorgiou for help with Python codes for analysing the Raman data.

\bibliography{ascn}

\end{document}